\documentclass[twocolumn,english]{aastex6}
\usepackage[T1]{fontenc}
\usepackage[latin9]{inputenc}
\usepackage{units}
\usepackage{graphicx}
\usepackage{wasysym}

\makeatletter
\slugcomment{}
\shorttitle{}
\shortauthors{}

\makeatother

\usepackage{babel}
\begin{document}
\global\long\def\la{\mathcal{L}}
\global\long\def\units#1{\mathrm{\,#1}}
\global\long\def\diff{\mathrm{d}}

\title{Capture of free-floating planets by planetary systems}

\author{Nadav Goulinski and Erez N. Ribak}

\affil{Physics Department, Technion - Israel Institute of Technology, Haifa
32000, Israel}
\begin{abstract}
Evidence of exoplanets with orbits that are misaligned with the spin
of the host star may suggest that not all bound planets were born
in the protoplanetary disk of their current planetary system. Observations

have shown that free-floating Jupiter-mass objects can exceed the
number of stars in our galaxy, implying that capture scenarios may
not be so rare. To address this issue, we construct a three-dimensional
simulation of a three-body scattering between a free-floating planet
and a star accompanied by a Jupiter-mass bound planet. We distinguish
between three different possible scattering outcomes, where the free-floating
planet may get captured after the interaction with the binary, remain
unbound, or ``kick-out'' the bound planet and replace it. The simulation
was performed for different masses of the free-floating planets and
stars, as well as different impact parameters, inclination angles
and approach velocities. The outcome statistics are used to construct
an analytical approximation of the cross section for capturing a free-floating
planet by fitting their dependence on the tested variables. The analytically
approximated cross section is used to predict the capture rate for
these kinds of objects, and to estimate that about 1\% of all stars
are expected to experience a temporary capture of a free-floating
planet during their lifetime. Finally, we propose additional physical
processes that may increase the capture statistics and whose contribution
should be considered in future simulations.
\end{abstract}


\section{Introduction}

The canonical planet formation mechanism is an in-situ planet formation
theory, dictating that planets form inside a protoplanetary disk.
If so, planetary systems would have to be the only source of free-floating
planets. An alternative suggestion adopts an ex-situ approach, where
these planetary-mass free-floating objects (hereafter, referred as
\textquotedbl{}free-floaters\textquotedbl{}) may form by gravitational
collapse of interstellar gas blobs.

It was suggested that high-speed gas blobs from the explosive death
of stars may form free-floaters by accretion of interstellar ambient
matter as they slow down, cool by radiation and collapse into a hot
Jupiter once their mass exceeds the Jeans mass \citep{dado2011misaligned}.
Such blobs and other filamentary structures are observed in large
numbers in nearby supernova remnants \citep{fesen2006expansion},
planetary nebulae \citep{o2002knots,matsuura2009firework} or star
formation regions, and are considered to be common in these stellar
stages. Some of these blobs were observed as dense conglomerations
of gas within H II regions, were they are referred to as globulettes
\citep{haworth2015theory}.

Measurements of the angle between the planetary orbital axis and the
stellar spin axis (spin-orbit angle) reveal that a considerable fraction
of the hot Jupiters have misaligned spin-orbit. This misalignment
does not settle with the expectation of a close alignment between
the spin of the star and the orbital motion of the planets, as they
all should inherit their angular momentum from the protostellar disk.
As for now, the spin-orbit angle of 87 planets was calculated from
light curves that exhibit anomalies due to the Rossiter-McLaughlin
effect. About 40\% of them show significant spin-orbit misalignment,
and nine of them are retrograde planets \citep{campante2016spin}.

Observations of supernova remnants and planetary nebulae each show
thousands of blobs that can lead to considerable number densities
of free-floaters, which may be large enough for captures to be common.
A wide-field image of the Helix Nebula in the $2.12\units{\mu m}$
molecular hydrogen line shows more then 40,000 blobs that constitute
the only source of the $\mathrm{H}_{2}$ surface brightness \citep{matsuura2009firework}.
If we assume that the only source of free-floaters are planetary nebulae
and that the $\sim40,000$ blobs found in the Helix Nebula are a typical
number for stars at this evolutionary stage, then a planetary nebulae
number density of $\sim5\times10^{-3}\units{pc^{-3}}$ \citep{holberg2002determination}
predicts a blobs number density of $n_{f}\approx200\units{pc^{-3}}$.
This tells us that there may be one thousand Jupiter mass objects
for every star in the Galaxy. In order to predict the capture rate
of this kind of free-floaters by planetary systems, we evaluate the
capture cross section by simulating three-dimensional scatterings
between a planetary-mass free-floater and a star-planet binary. \\

\section{Approach}

Significant orbit perturbations of a bound planet due to interactions
with free-floaters are expected to occur for impact parameters of
$-b_{max}\leq b\leq b_{max}$, where the value of $b_{max}$ is determined
by evaluating the maximal closest approach distance $r_{min}$ from
the host star, for which significant orbital perturbations are still
possible. Let us assume that a free-floater with a relative velocity
$v_{\infty}$ at infinity is approaching a star-planet binary with
an impact parameter $b$. Assuming that the mass of the star $M$
is much larger then the mass of the bound planet $m_{B}$ and the
free-floater $m_{f}$, the trajectory of the free-floater is the one
of a test-particle, and determined by the mass of the star. The cross
section for significant perturbations induced by this flux is the
product of the geometrical area times the gravitational focusing factor
\begin{eqnarray}
\mathcal{B} & = & \pi b^{2}=\pi r_{min}^{2}\left(1+\frac{2GM}{v_{\infty}^{2}r_{min}}\right).\label{eq: Geometrical}
\end{eqnarray}
We use an analytical expression for the closest approach $r_{min}$,
that holds for $m_{f},m_{B}\ll M$ \citep{donnison1984stability}
\begin{equation}
r_{min}=r_{B}\frac{\left(\left(1+\rho\right)^{\nicefrac{3}{2}}-1\right)^{2}}{2\rho^{2}},\label{eq:r_min}
\end{equation}
where $r_{B}$ is the semi-major axis of the bound planet, and $\rho$
is the the mass ratio between the free-floater and the bound planet
(hereafter, the planetary-mass ratio).

We may write the differential size of the capture cross section $\diff\sigma$
as a product of the differential cross section for a significant interactions
$2\pi b\diff b$, times the fraction of interactions that resulted
in a capture $P$
\begin{equation}
\diff\sigma=P\cdot2\pi b\diff b.
\end{equation}
The interaction cross section depends on the relative velocity $v$,
the mass of the host star $M$, the planetary-mass ratio $\rho$,
and the semi-major axis of the bound planet $r_{B}$; the last two
determine the closest approach radius. The capture probability may
also depend on similar variables, but dependence on relative velocity
and the inclination angle is expected, since the perturbing impulse
is affected.

Let us assume that a homogeneous flux of free-floaters with a given
mass $m_{f}$, relative velocity $v$ and a number density $n_{f}$
is approaching with a given inclination angle $\theta_{i}$ at a planetary
system. This flux will produce $n_{f}v\diff\sigma\left(v,\theta,M,r_{min}\right)/\diff\Omega$
captures per second per solid angle. For a given relative velocity
dispersion $\mathcal{S}$, the capture rate for a specific stellar
mass and $r_{min}$ is obtained by integrating with respect to velocity
and over the full solid angle 
\begin{eqnarray}
\mathrm{R}\left(M,r_{min}\right) & = & \iint n_{f}v\frac{\mathrm{d}\sigma\left(v,\theta,M,r_{min}\right)}{\mathrm{d}\Omega}f_{\mathcal{S}}\diff v\diff\Omega,\label{eq:Spesific Rate}
\end{eqnarray}
where $f_{\mathcal{S}}$ is the relative velocity distribution. The
expected capture rate is obtained after integrating with respect to
the $M$ and $r_{min}$ according to their corresponding distribution
function, and multiplying by the total number of stars $N_{stars}$
\begin{equation}
\left\langle \mathrm{R}\right\rangle =N_{stars}\int\mathrm{R}\left(M,r_{min}\right)f\left(M,r_{min}\right)\mathrm{d}Mdr_{min}.\label{eq:Total Rate}
\end{equation}

\section{\label{sec:Numerical-Method}Numerical Method }

We follow the work of \citet{varvoglis2012interaction} where they
simulated scattering events of a Jupiter-mass free-floater by a binary
of Sun-mass star and a Jupiter-mass planet. The free-floater was placed
on a parabolic coplanar orbit and sufficiently far away to be considered
as an unbound object. They mapped the resulting outcomes of the simulation
for a grid of impact parameters $b$ and initial orbital phases $\phi_{B}$
of the planetary system, and calculated the fraction of captures $P$.
This is because different initial orbital phases lead to different
impulse durations and strengths. It turned out that about $50\%$
of the initial condition grid led to a so-called ``temporary capture'',
where the fraction of captures that end up with moderate values of
semi-major axis and eccentricity is in the order of $1\%$. These
captures are considered as temporary because simulating this isolated
system for a large number of revolutions of the captured free-floater
shows that most of them will eventually gain energy and escape. They
also performed simulations for different free-floater masses, showing
that the fraction of captures grows linearly with the mass, but not
significantly.

\begin{figure}[h]
\begin{centering}
\includegraphics[width=\columnwidth]{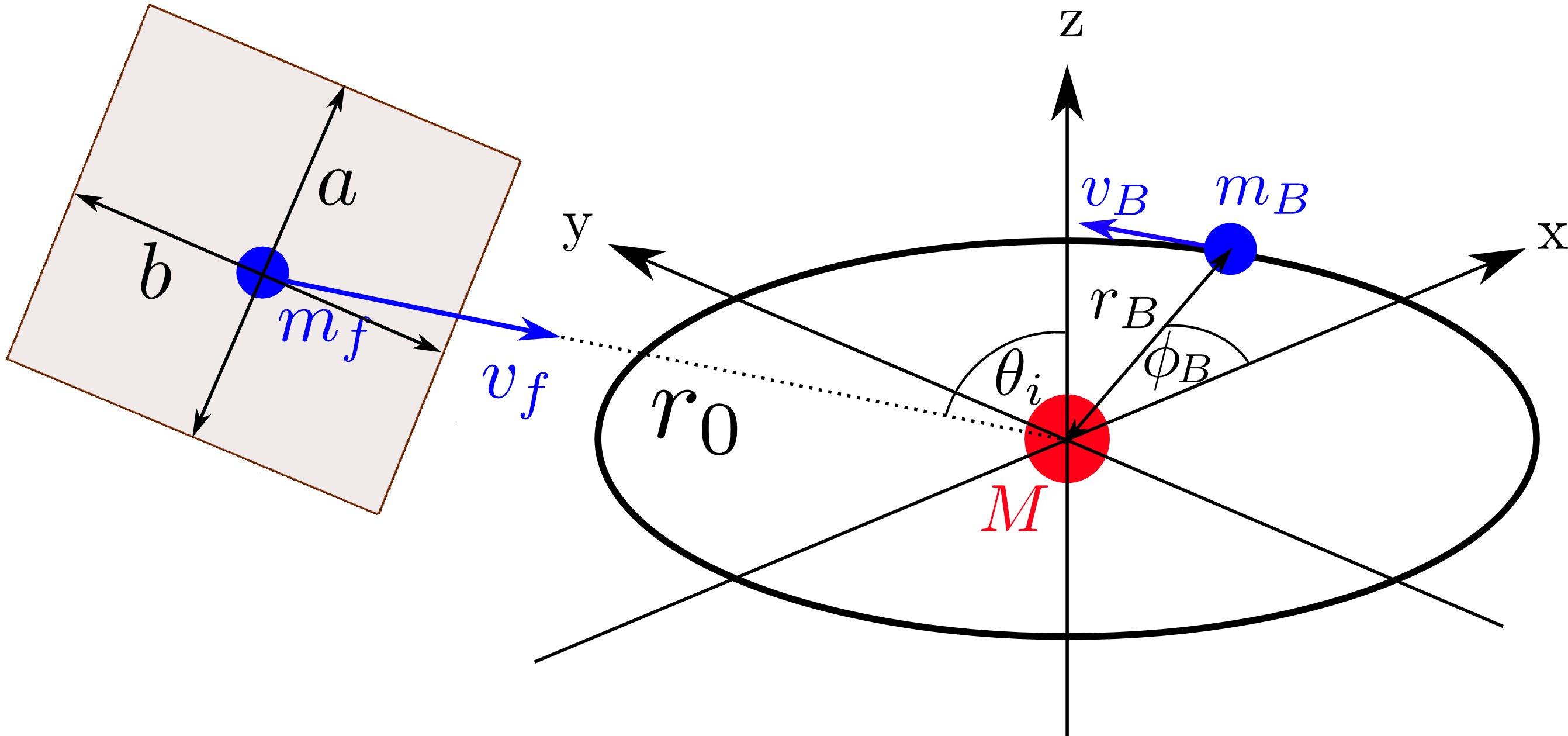}
\par\end{centering}
\caption{{\footnotesize{}Schematic diagram of the scattering simulation. A
free-floater (blue with the mass $m_{f}$) approaches a star-planet
binary with a velocity of $v_{f}$ and an inclination angle $\theta_{i}$.
The brown plane, which is orthogonal to the velocity vector, depicts
the two components of the impact parameter $d=\sqrt{b^{2}+a^{2}}$.
Cases of $a\protect\neq0$ will be tested separately. The bound planet
(blue with the mass $m_{B}$) revolves around the star (red with the
mass $M$) on a circular orbit, with a radius of $r_{B}$ and velocity
$v_{B}=\dot{\phi}_{B}r_{0}$. The free-floater is placed at a distance
of $r_{0}\gg r_{B}$ from the $y$ axis. }}
\label{fig:Schematic-diagram-of}
\end{figure}

The free-floater, as well as the host star and the bound planet, are
set as point-mass objects. The mass of the star and the free-floater
are simulation variables, while the mass of the bound planet is of
one Jupiter-mass. The bound planet is set to revolve around the host
star on simple circular orbit. We define the inclination angle to
be the one between the initial velocity vector of the free-floater
$v_{f}$ and the orbital axis of the bound planet, (see Fig.~\ref{fig:Schematic-diagram-of}).
The initial distance of the free-floater from the system is set to
be $r_{0}=\sqrt{40m_{f}/m_{B}r_{B}+b^{2}}$, so that the initial binding
energy between the free-floater and the binary is negligible. The
corresponding initial velocity is obtained through conservation of
energy 
\begin{equation}
\frac{1}{2}m_{f}v_{\infty}^{2}=\frac{1}{2}m_{f}v^{2}-\frac{Gm_{f}(M+m_{B})}{\sqrt{b^{2}+\left(40r_{B}\right)^{2}}}.\label{eq:Unbound}
\end{equation}
 
The simulations are carried for different masses of the free-floater
and star, velocities and inclination angles. Each simulation run preforms
multiple scattering that corresponds to a grid of impact parameters
$b$ and initial orbital phases of the bound planets $\phi{}_{B}$.
The array of impact parameters $b$ spans up to some maximal value
$b_{max}$, for which significant perturbation may still occur (Eq.~\ref{eq:r_min}).

One may notice from Fig.~\ref{fig:Schematic-diagram-of} that all
values of $b$ lead for the path of the free-floater to intersect
with the y-axis if no interaction with the bound planet took place.
This means that the capture statistics that obtained from a simulation
for some inclination angle $\theta_{i}$ do not necessarily predict
the percentage of captures produced by a flux of free-floaters that
passes through an area of $\pi b^{2}$. To account for this, the effect
of a non-zero second component of the impact parameter, orthogonal
to $b$ and the velocity vector $v_{0}$, will be be tested separately.

The dynamical system is described by a simple Lagrangian. It includes
the kinetic term of the two planets, the interaction term between
them, and their interaction terms with the star
\begin{eqnarray*}
\la_{kinetic} & = & \frac{1}{2}m_{f}v_{f}^{2}+\frac{1}{2}m_{B}v_{B}^{2}\\
\la_{interaction} & = & G\frac{Mm_{f}}{r_{f}}+G\frac{Mm_{B}}{r_{B}}+G\frac{m_{f}m_{B}}{\left|\mathbf{r}_{f}-\mathbf{r}_{B}\right|}.
\end{eqnarray*}
We do not add a kinetic term for the star since we assume standard
planetary masses, for which the displacement of the star due to interaction
is negligible down to red dwarf masses. 

\begin{figure}[h]
\begin{centering}
\includegraphics[width=10.5pc]{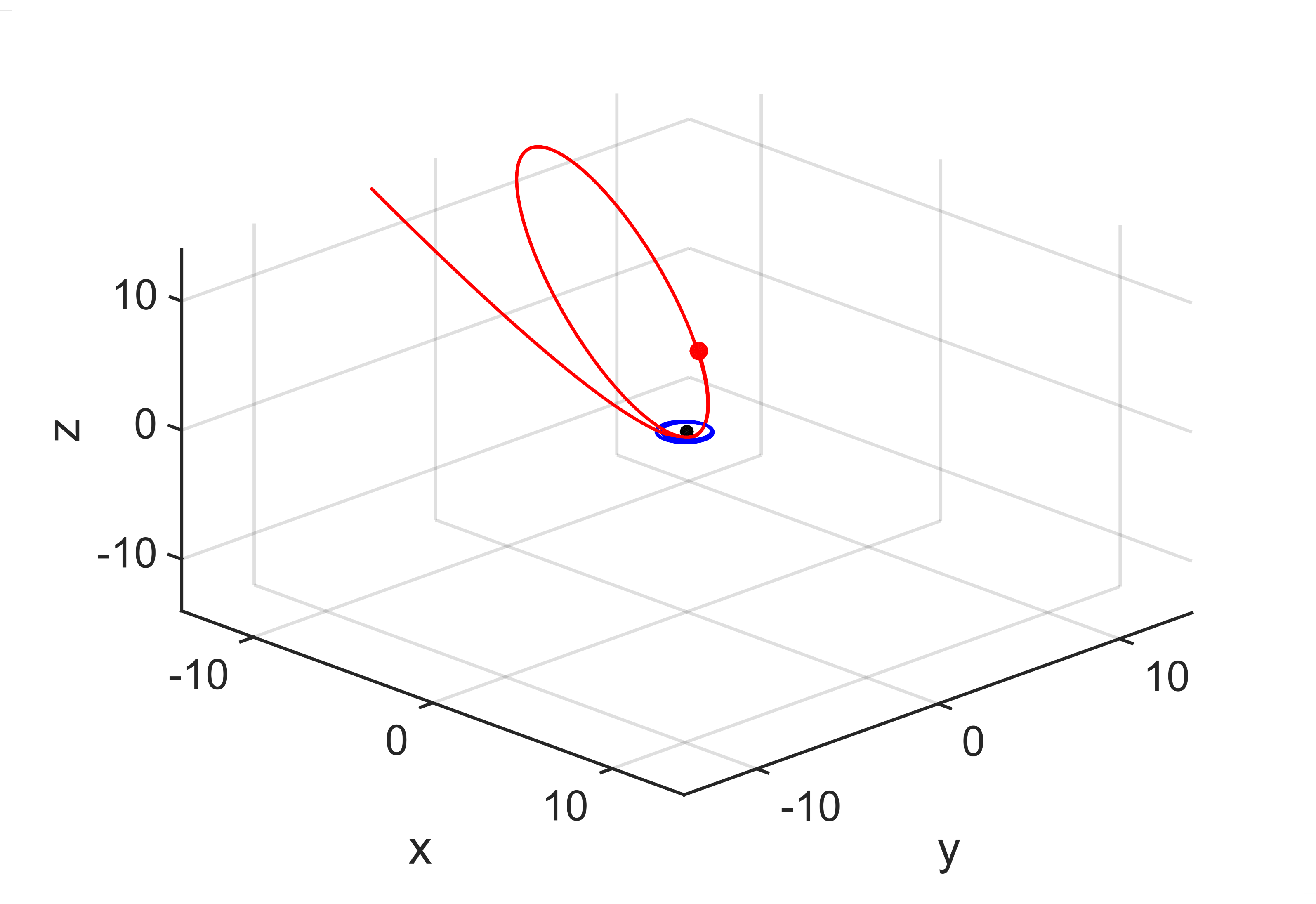}\includegraphics[width=10.5pc]{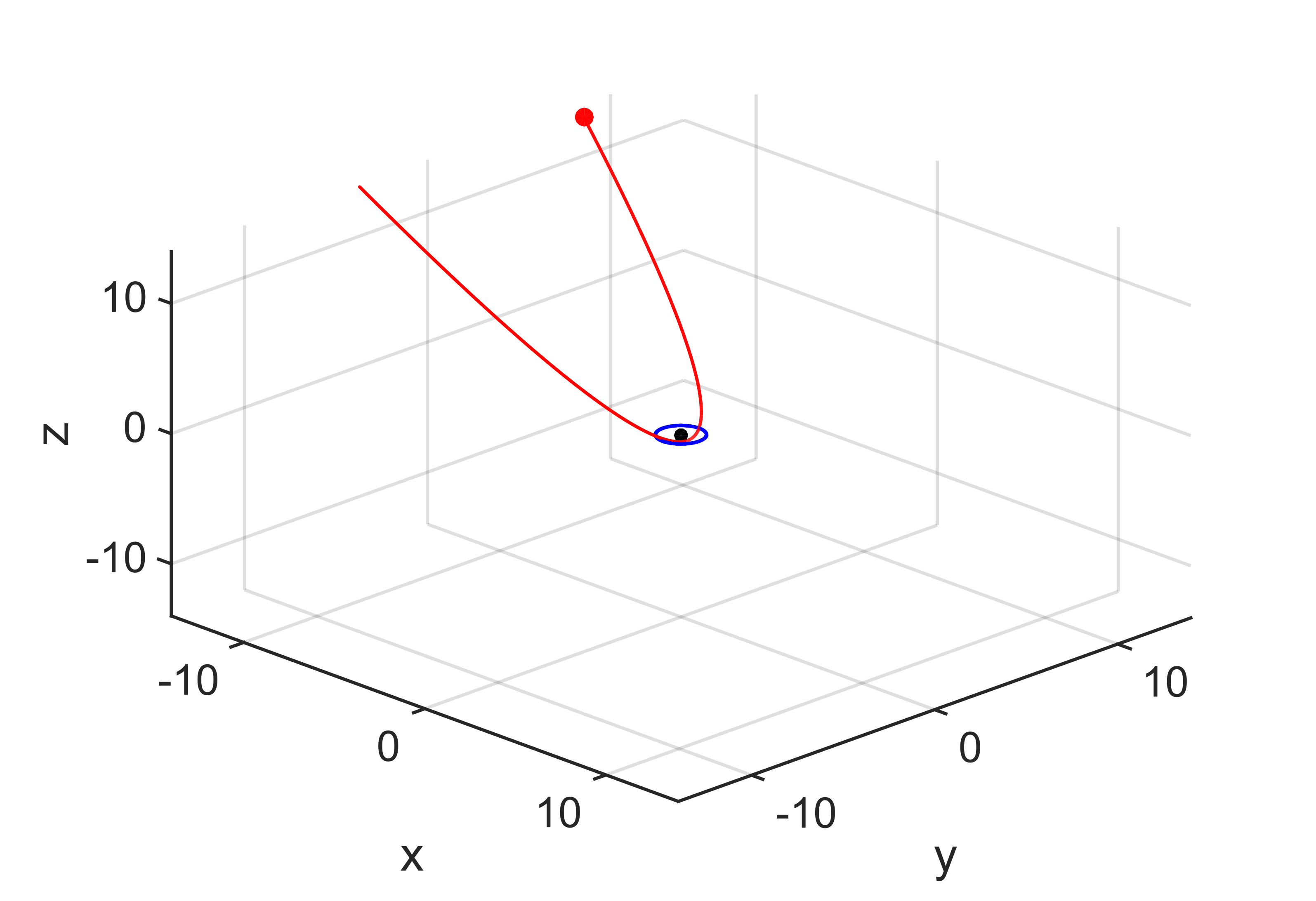}\\
\includegraphics[width=10.5pc]{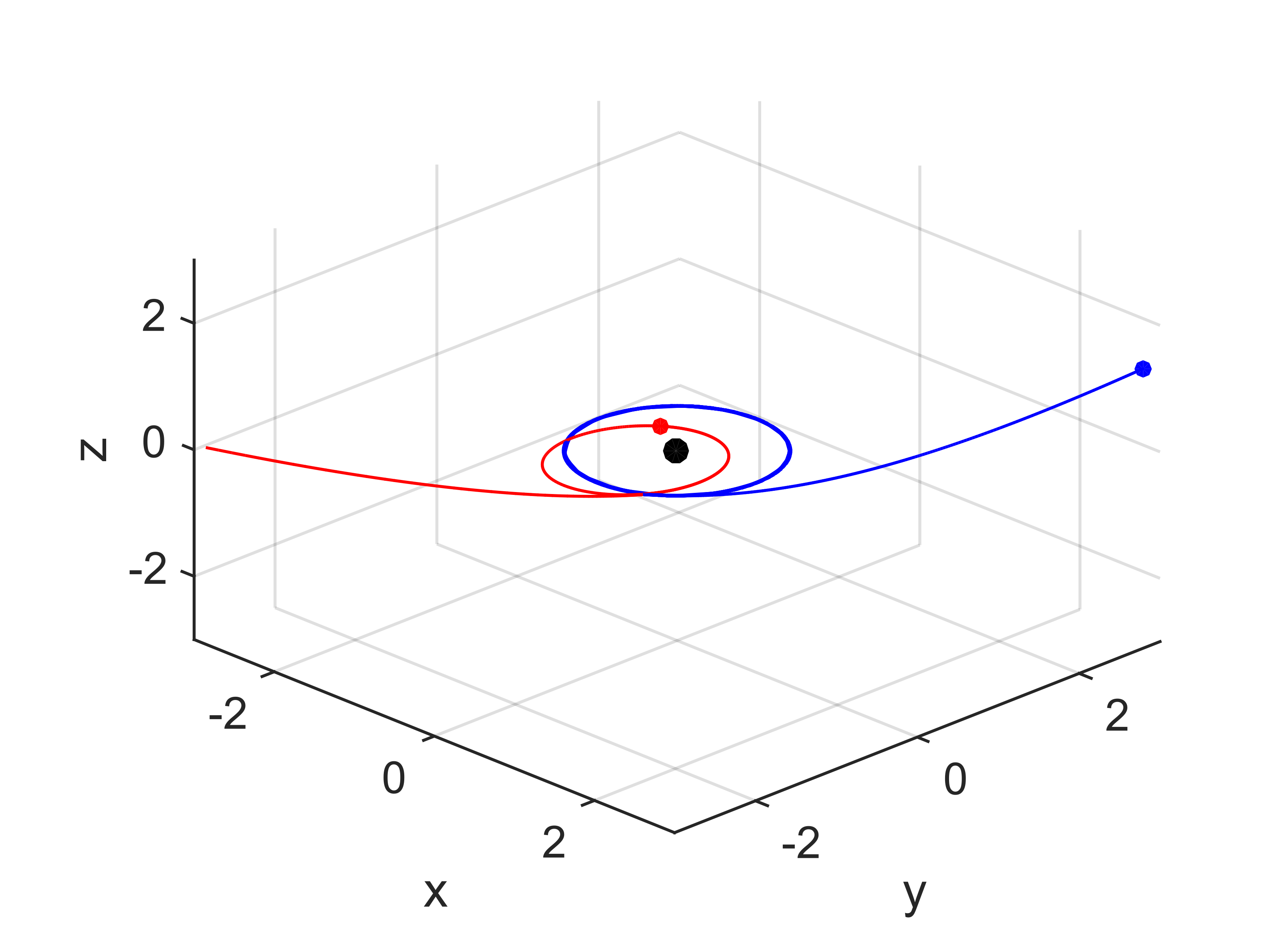}
\par\end{centering}
\caption{{\footnotesize{}The interaction between the free-floater and the bound
planet leads to three different possible outcomes of interest - a
capture, a flyby and an exchange, where the free-floater ``replaces''
the bound planet; these are visualized above in a clockwise order
from top left.}}
{\footnotesize{}\label{fig:different_outcomes}}{\footnotesize \par}
\end{figure}

We preform the simulation for 175 values of $-b_{max}<b<b_{max}$
and 125 values of $0<\phi<2\pi$ for each inclination angle, velocity
at infinity, mass of the free-floater and mass of the host star. The
equations of motion are integrated until the free-floater travels
out approximately twice its initial distance from the star. At this
time, we calculate the sum of the kinetic and potential energy for
two pairs: $E_{f}$ - star and free-floater, and $E_{B}$ - star and
bound planet. The sign of the energy distinguishes between three outcomes,
visualized in Fig.~\ref{fig:different_outcomes}:
\begin{itemize}
\item Flyby: $E_{f}\geq0,~E_{B}<0$
\item Capture: $E_{f}<0,~E_{B}<0$
\item Exchange: $E_{f}<0,~E_{B}\geq0$
\end{itemize}
The outcomes are saved together with their corresponding grid parameters
$\left(b,\phi_{B}\right)$, for which the fraction of captures $P$
is calculated.

In order to maximize the simulation speed, we used a grid resolution
that is four times sparser then the one used by \citet{varvoglis2012interaction},
but still high enough for statistical purposes. The numerical precision,
however, must be sufficient enough so that the error in energy will
be at least one order of magnitude smaller than the value itself.
The numerical precision is set by two parameters: the first is the
relative tolerance $r_{T}$, which determines the maximal relative
error of the solution, and the second is the absolute tolerance $a_{T}$,
which determines the last important digit of the solution. These parameters
were set to $r_{T}=a_{T}=10^{-7}$, for which the accuracy and the
simulation speed are sufficient. We assume that the maximal error
in every solution component $y_{i}$, taken as $\left|e_{i}\right|=r_{T}\left|y_{i}\right|+a_{T}$,
has been achieved, so that the maximal error in $E_{f}$ is
\[
\delta E_{f}=\sqrt{\sum_{i}\left(\frac{\partial E_{f}}{\partial y_{i}}\left|e_{i}\right|\right)^{2}},
\]
where $E_{f}=\nicefrac{1}{2}m_{f}\dot{\mathbf{r}}^{2}-m_{f}/r$. We
calculate $\delta E_{f}$ for every run of the simulation to assure
that the number of runs that resulted with $\delta E_{f}/E_{f}>0.1$
is less then $1\%$.

\section{Results}

\begin{figure}[h]
\begin{centering}
\includegraphics[width=1\columnwidth]{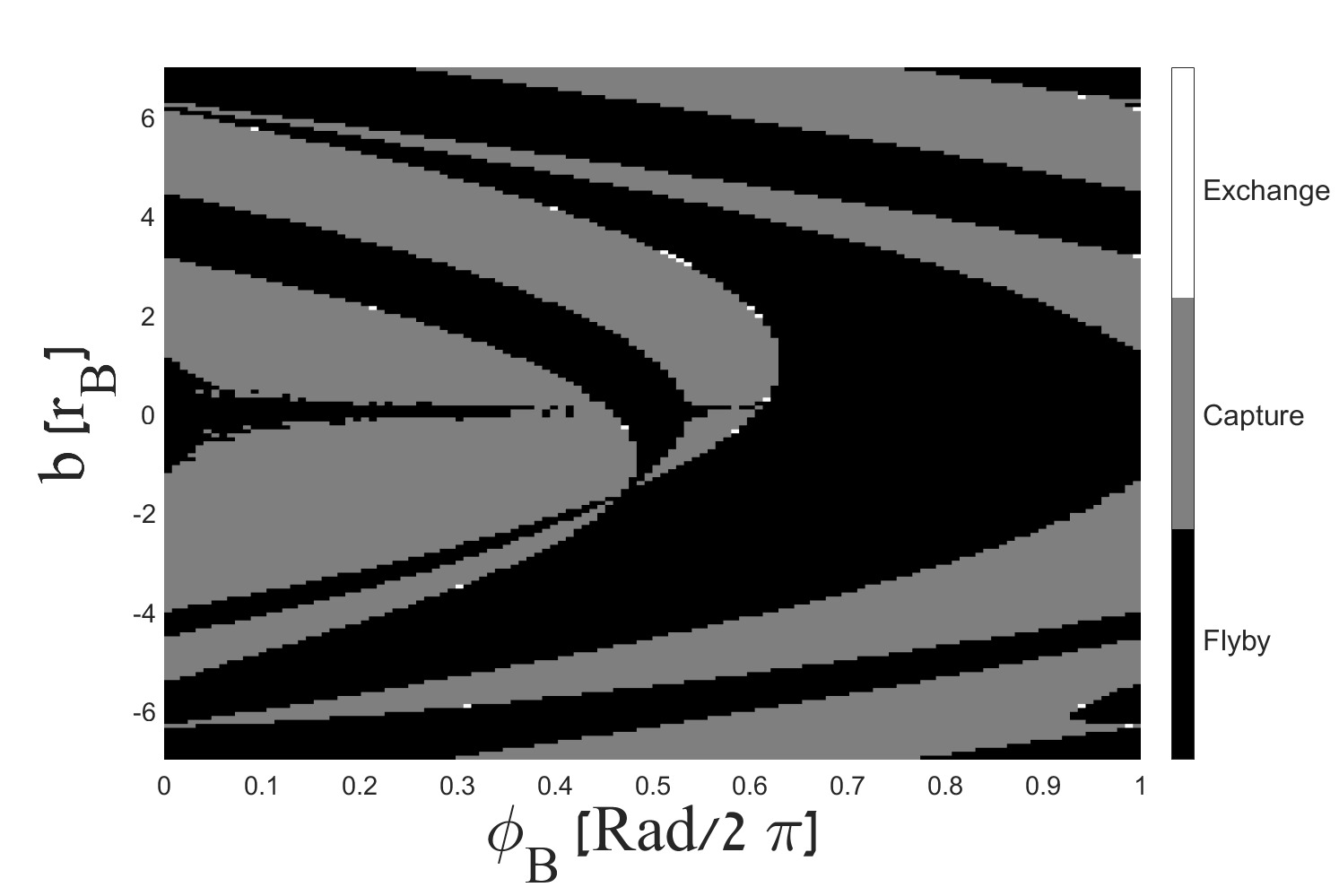}
\par\end{centering}
\caption{{\footnotesize{}An outcome map for an inclination angle of $\theta_{i}=90^{\circ}$,
where the scatterings between the Jupiter-mass free-floater and a
star-planet system are coplanar. Each point corresponds to a specific
impact parameter $b$ (175 in total) and an initial orbital phase
of the bound planet $\phi{}_{B}$ (125 in total). An outcome where
the free-floater remains unbound (flyby) is represented by a black
color, while the case where the free-floater gets captured is represented
by a gray color. Special cases where the free-floater ``kicks out''
the bound planet and replaces it are seen as white dots on the boundary
between the two main outcomes. The resulting capture probability is
$P=48.57\%$.}}
{\footnotesize{}\label{fig:An-outcome-map 175x125}}{\footnotesize \par}
\end{figure}

We start with a Jupiter-mass free-floater that approaches the binary
on a parabolic and coplanar orbit $\left(v_{\infty}=0,\theta_{i}=90^{\circ}\right)$.
The resulting outcome map is displayed in Fig.~\ref{fig:An-outcome-map 175x125},
where we use gray levels to distinguish between captures, flybys,
and exchanges. The gray area, which represents regions of $b,\phi{}_{B}$
values that led to captures, covers almost $50\%$ of grid. The upper
an lower impact parameter values that were tested are those beyond
which exchanges do not occur, indicating that significant orbital
perturbations are no longer possible or too rare. Although statistically
negligible, exchange events may also produce misaligned orbits, and
they are considered as captures.
 This outcome map agrees with the one presented by \citet{varvoglis2012interaction},
implying the consistency of the method. Additionally, a variety of
$b,\phi{}_{B}$ grid values with specific outcomes were selected from
the map and tested individually to assure that the final energy values
do not vary, and that the trajectories of the bodies agree with the
outcomes. 

\begin{figure}[h]
\begin{centering}
\includegraphics[width=1\columnwidth]{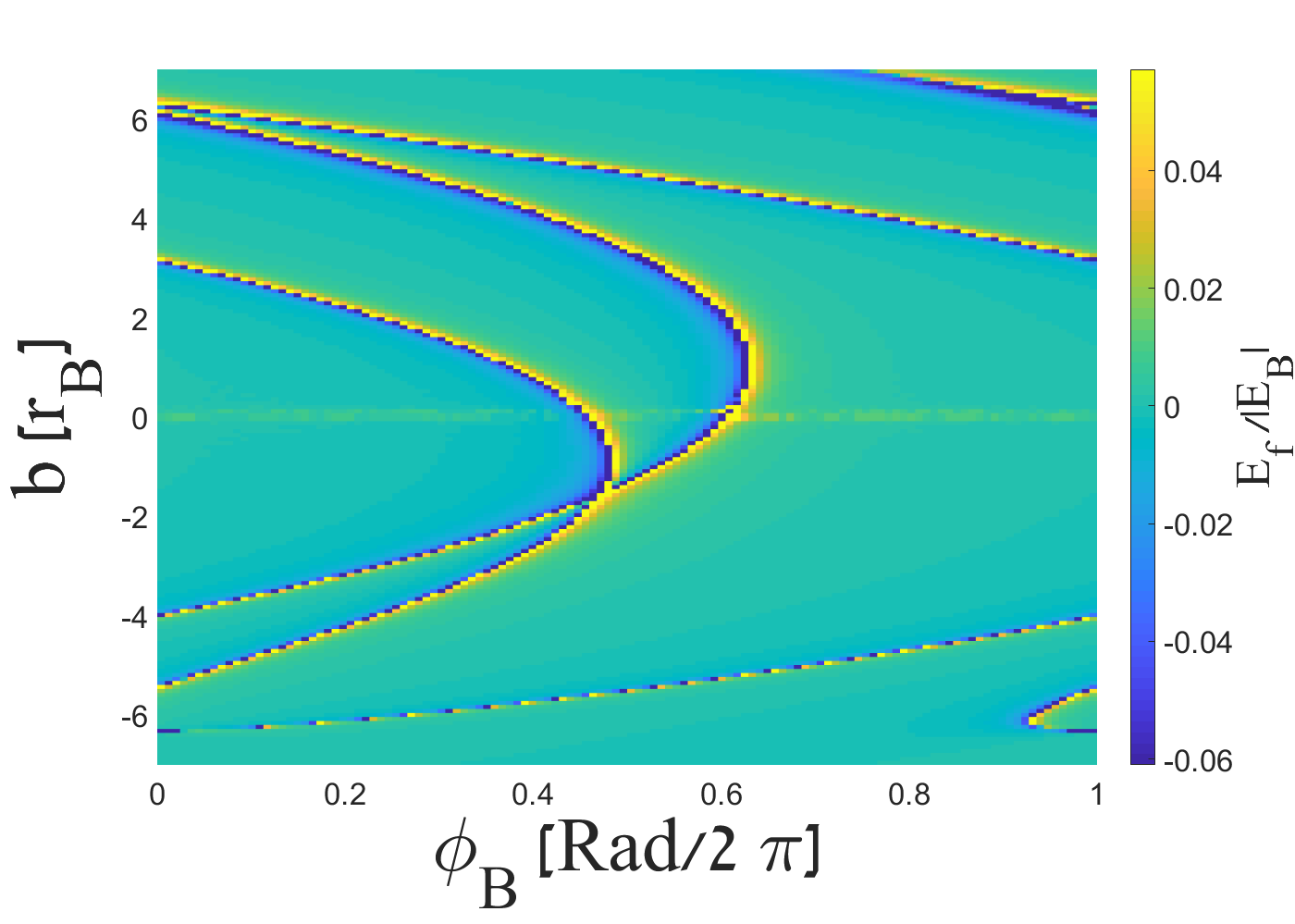} 
\par\end{centering}
\caption{{\footnotesize{}The final energy map of the free-floater, corresponding
to }Fig.~{\footnotesize{}\ref{fig:An-outcome-map 175x125}. The color
spectrum represent the ratio between the final energy of the free-floater
and the initial energy of the bound planet. The turquoise color that
covers most of the map represents energy ratio values which are very
close to zero (the free-floater's original orbit is almost unaffected)
while the yellow color indicates flyby events with a significant excess
of energy and blue color indicate capture events with a significant
negative energy (tightly bound). Values above 0.25 or below -0.25
were cropped. }}

\label{fig:Energy-map 700x500}
\end{figure}

The majority of the captures, however, end up with energies that are
$\sim10^{-3}$ time smaller then the initial energy of the binary,
and are very eccentric and elongated. The outcome of the simulations
is also plotted in Fig.~\ref{fig:Energy-map 700x500} in terms of
the ratio between the final energy of the free-floater and the initial
energy of the binary. One can see that the maximal energy transfer
takes place around the boundary between capture and flyby regions,
at which exchange events are prone to happen. For most $b,\phi{}_{B}$
values, the total energy of the free-floater is almost unchanged.
There is ``line'' of significant positive energies around $b=0$,
where the minimal approach of the free-floater from the star is comparable
to the numerical precision, and deviation from conservation of energy
becomes substantial. This small region is disregarded in the statistics.

\begin{figure}[h]
\begin{centering}
\includegraphics[width=1\columnwidth]{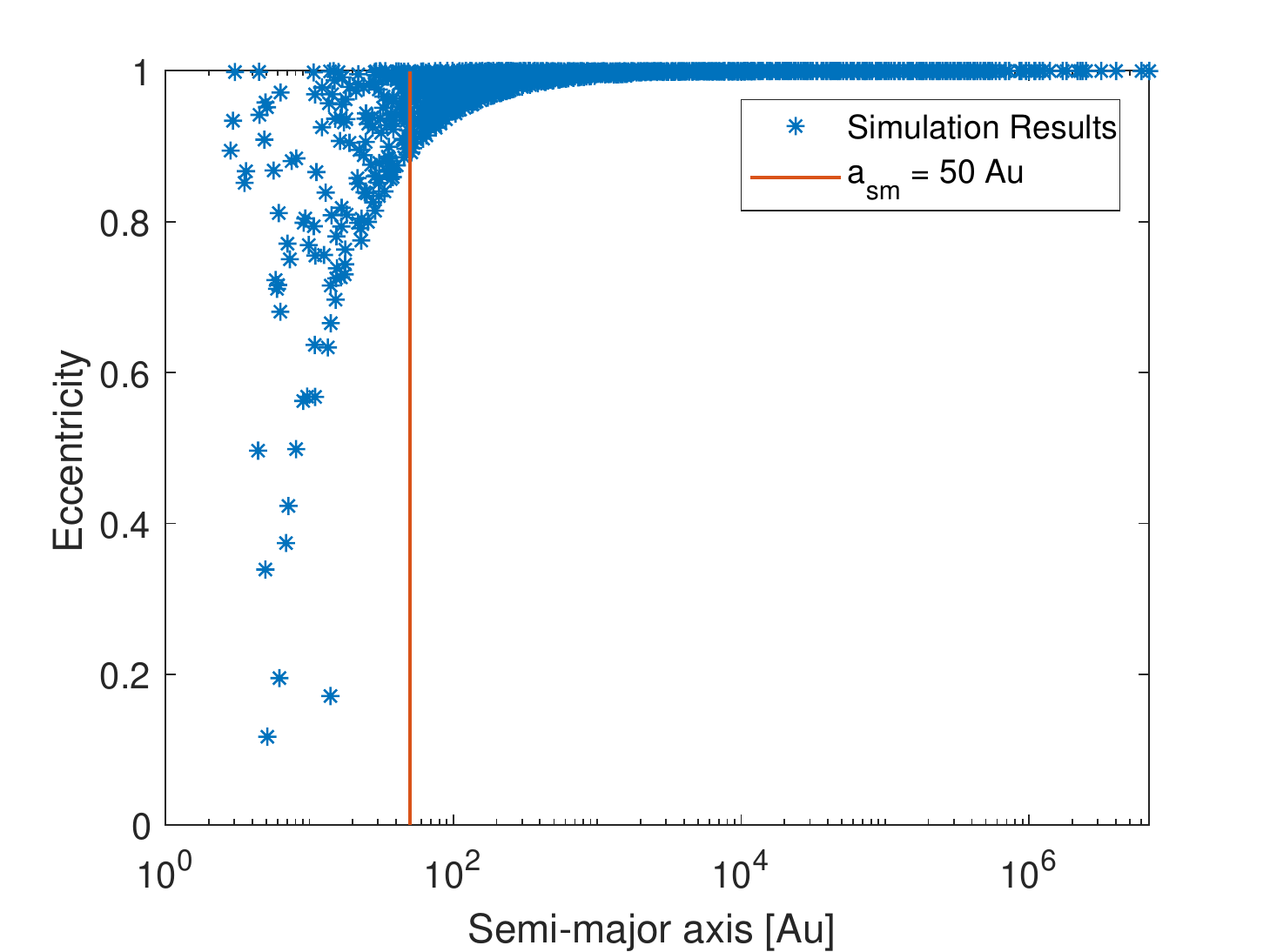}
\par\end{centering}
\caption{{\footnotesize{}Eccentricity of the captured free-floaters as a function
of the semi-major axis for a coplanar configuration and zero relative
velocity at infinity. Only $\sim0.9$\% of the simulated captures
ended up with a semi-major axis of $a_{sm}<50\protect\units{AU}$,
which is marked by the vertical red line.}}
{\footnotesize{}\label{fig:Eccentricity edge-on}}{\footnotesize \par}
\end{figure}

The eccentricity as a function of the semi-major axis for the captured
Jupiter-mass free-floaters is given in Fig.~\ref{fig:Eccentricity edge-on}.
 Only $\sim0.9$\% of the simulated scattering events ended up with
a semi-major axis of $a_{sm}<50\units{AU}$, which is the distance
of the Kuiper belt from the sun. According to our result, almost all
free-floaters will reach their aphelion beyond the Kuiper belt if
they would be captured by our Sun, but the number of $a_{sm}<50\units{AU}$
captures drops rapidly for slight deviations from coplanarity. 

\begin{figure}[h]
\begin{centering}
\includegraphics[width=\columnwidth]{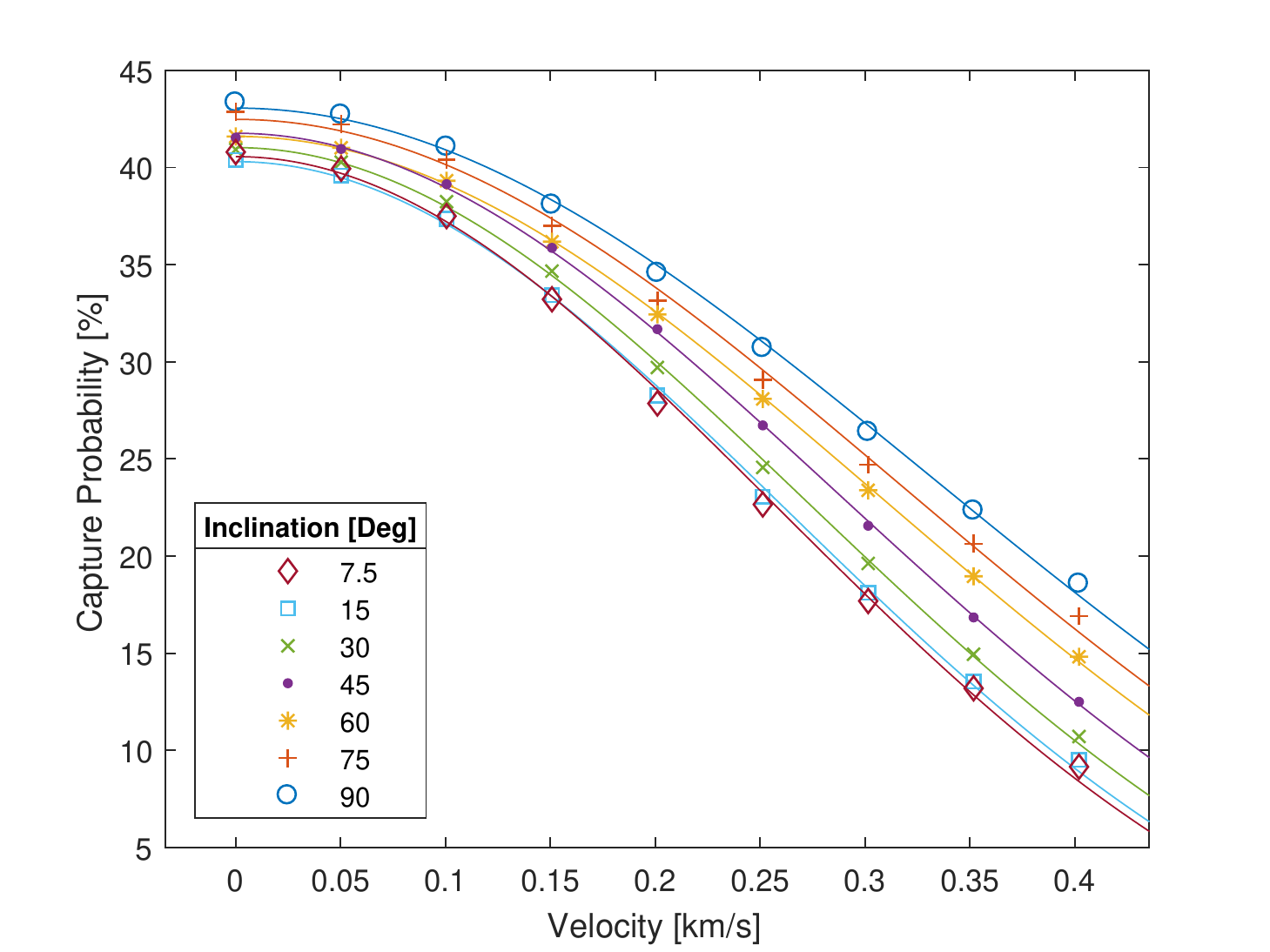}
\par\end{centering}
\caption{{\footnotesize{}The dependence of the capture probability on the initial
velocity at infinity of the free-floater. This dependence is plotted
for different inclination angles of the star-planet system. The continuous
curves represent the fitted Gaussian profiles, $P\left(v\right)=P_{0}\exp\left(-v^{2}/2\Sigma^{2}\right)$.}}
{\footnotesize{}\label{fig:Probability_Vs_Velocity_all_inc-1}}{\footnotesize \par}
\end{figure}

We cover a $v_{\infty}$ initial velocity range for which the capture
probability experiences significant variations. To evaluate the effect
of the lateral impact parameter $a\neq0$ on the statistics (Fig.~\ref{eq: Geometrical}),
we ran additional simulations for a Jupiter mass free-floater with
a fixed maximal value of $a=b_{max}$. The resulting statistics do
not show dramatic variations in terms of capture probability, and
they are averaged with the ones for $a=b_{max}$. Fig.~\ref{fig:Probability_Vs_Velocity_all_inc-1}
shows the dependence of the averaged capture probability on the velocity
of the free-floater for all different inclination angels that were
tested.
 The capture percentage is lower for free-floaters with higher initial
velocity and lower inclination angle (towards a face-on inclination),
which is expected since the orbital perturbation of the bound planet
gets shorter. The resulting functionality is best fitted with a Gaussian
$P\left(v\right)=P_{0}\exp\left(-v^{2}/2\Sigma^{2}\right)$, where
the parameter $P_{0}$ is capture fraction for the case of $v_{\infty}=0$
and $\Sigma$ is the standard deviation of the Gaussian profile.

\begin{figure}[h]
\begin{centering}
\includegraphics[width=\columnwidth]{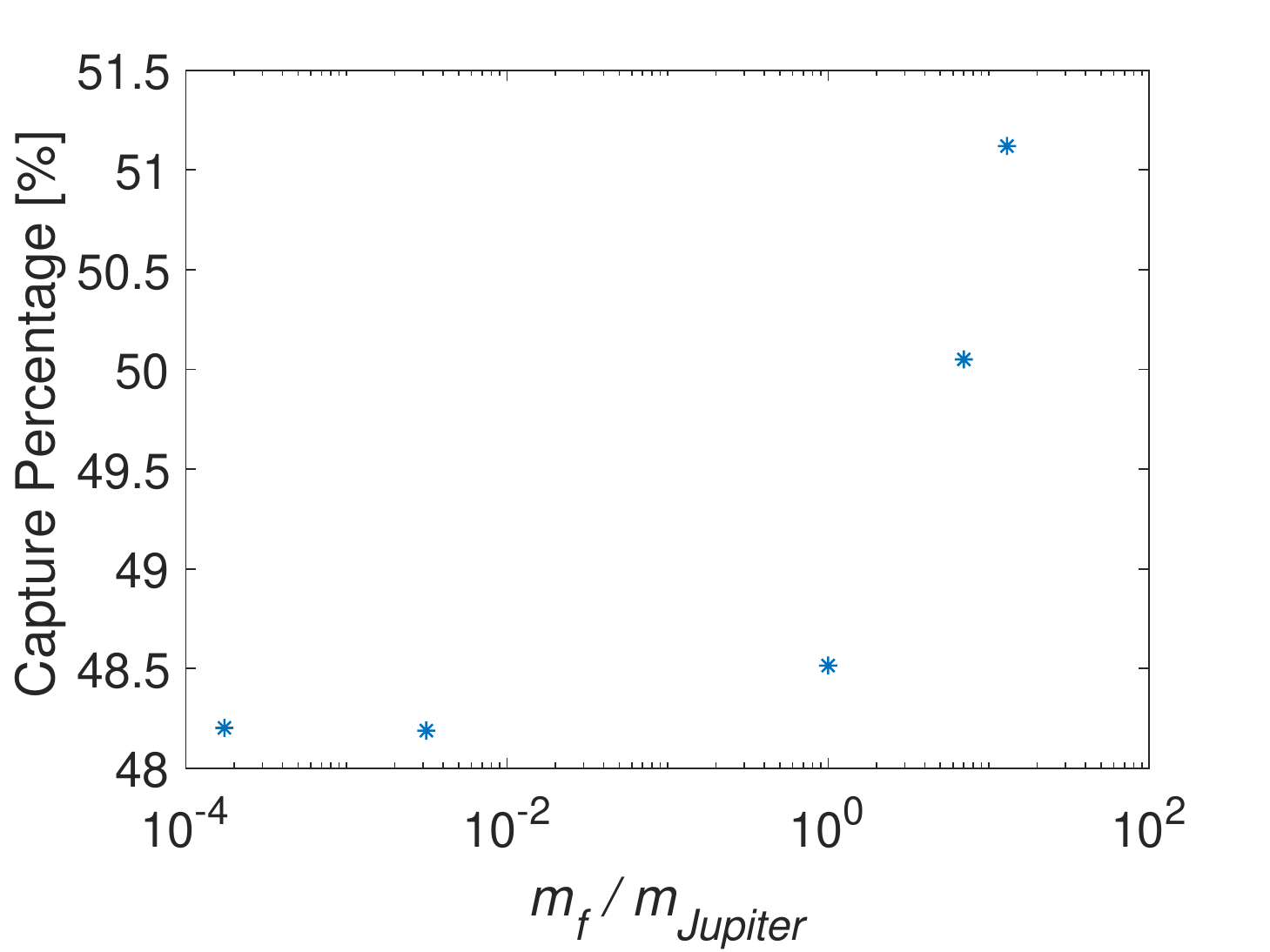}
\par\end{centering}
\caption{{\footnotesize{}Capture percentage obtained from simulations of five
different free-floater masses approaching the star-planet system at
a coplanar configuration and zero initial velocity at infinity. The
capture percentages obtained for for $m_{f}=7m_{\jupiter}$ and $m_{f}=13m_{\jupiter}$
are slightly higher, but are still around $50\%$.}}
{\footnotesize{}\label{fig:Capture-percentage-All-Masses}}{\footnotesize \par}
\end{figure}

To evaluate the dependence of the capture probability on the planetary-mass
ratio, we preform a full simulation for two additional lighter free-floaters
- Earth mass and Mercury mass and two additional coplanar scatterings
for $7m_{\jupiter}$ and $13m_{\jupiter}$ free-floaters. The statistics
for the different inclination angles and velocities hardly differed
from the ones for Jupiter-mass, and are effectively independent of
the mass at this regime. The later two exhibited a slight increase
in the capture percentage, as displayed in Fig.~\ref{fig:Capture-percentage-All-Masses}.

\begin{figure}[h]
\begin{centering}
\includegraphics[width=\columnwidth]{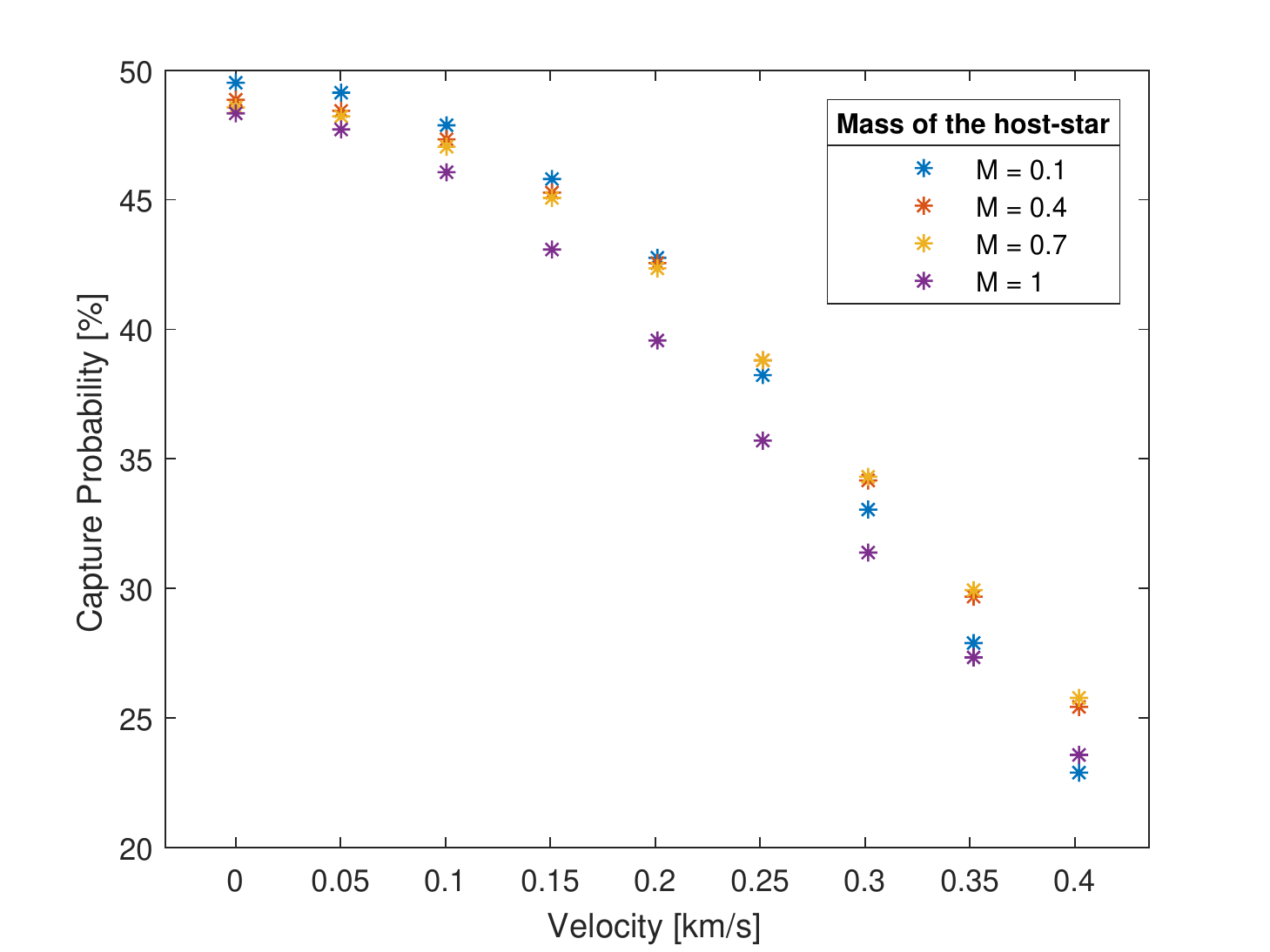}
\par\end{centering}
\caption{{\footnotesize{}\label{fig:StellarMass}The dependence of the capture
probability on the velocity of the free-floater at infinity, resulting
from coplanar scattering simulations. Different colors represent different
masses of the host-star.}}
\end{figure}

The dependence of capture probability on the stellar mass is also
needed to be analytically approximated. However, since 85\% of all
stars are sub-solar, we performed additional coplanar scattering simulations
with lower mass host-stars to evaluate their effect. The the resulting
capture probabilities are shown in Fig.~\ref{fig:StellarMass}, and
small differences of velocity dependence between the four tested masses
can be noticed. 
 This trend can be attributed to the shorter impulse that is needed
for the free-floater to loose its energy. Still, assuming that this
velocity dependence is best fitted with a Gaussian, the resulting
standard deviations differ only by $\sim0.3\units{km\cdot s^{-1}}$.
Since the typical velocity dispersion in our galaxy is in the order
of $\sim10^{2}\units{km\cdot s^{-1}}$, we do not expect the capture
rate to be significantly affected.

\section{The cross section construction}

\begin{figure}[h]
\begin{centering}
\includegraphics[scale=0.3]{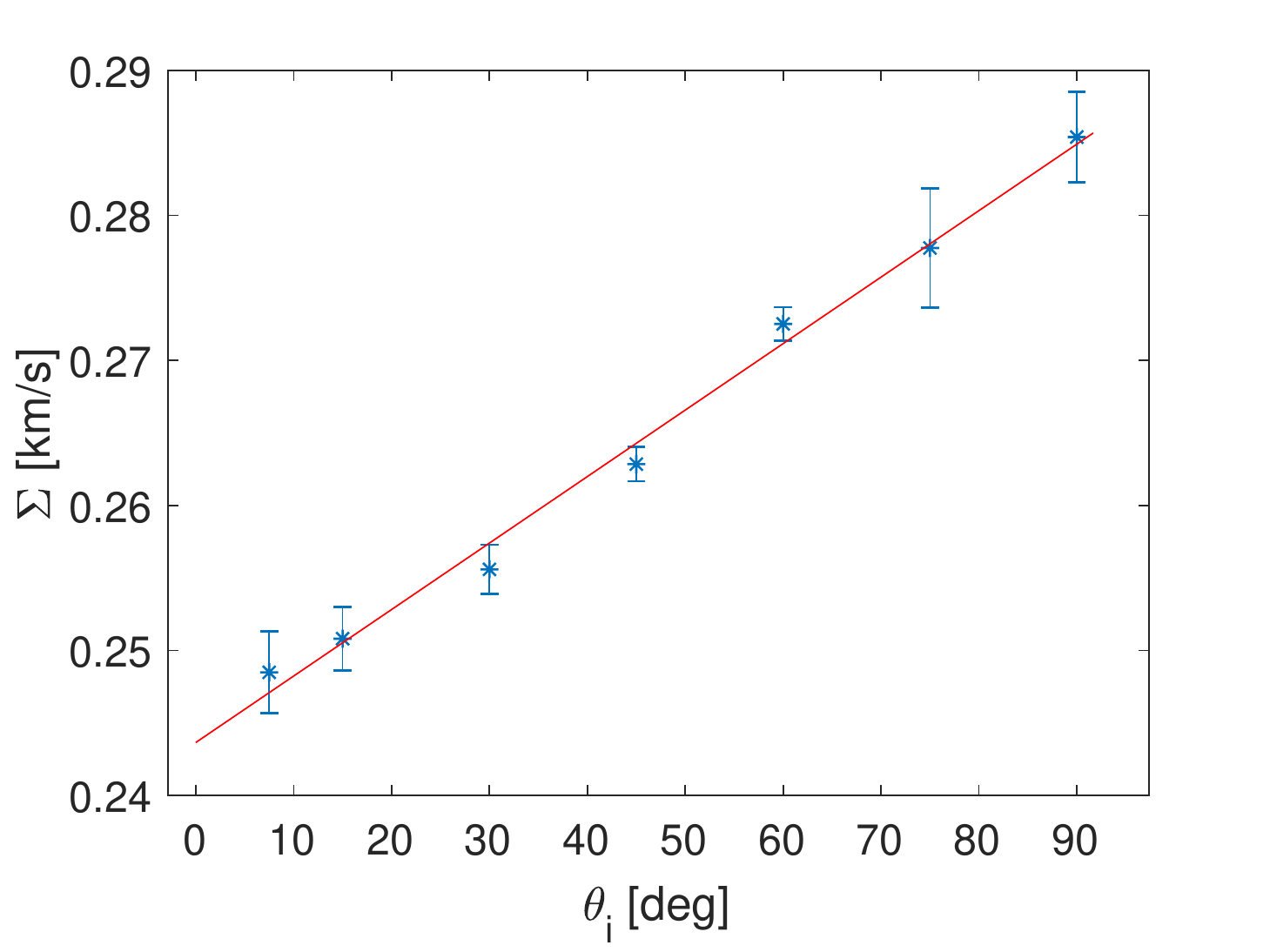}\includegraphics[scale=0.3]{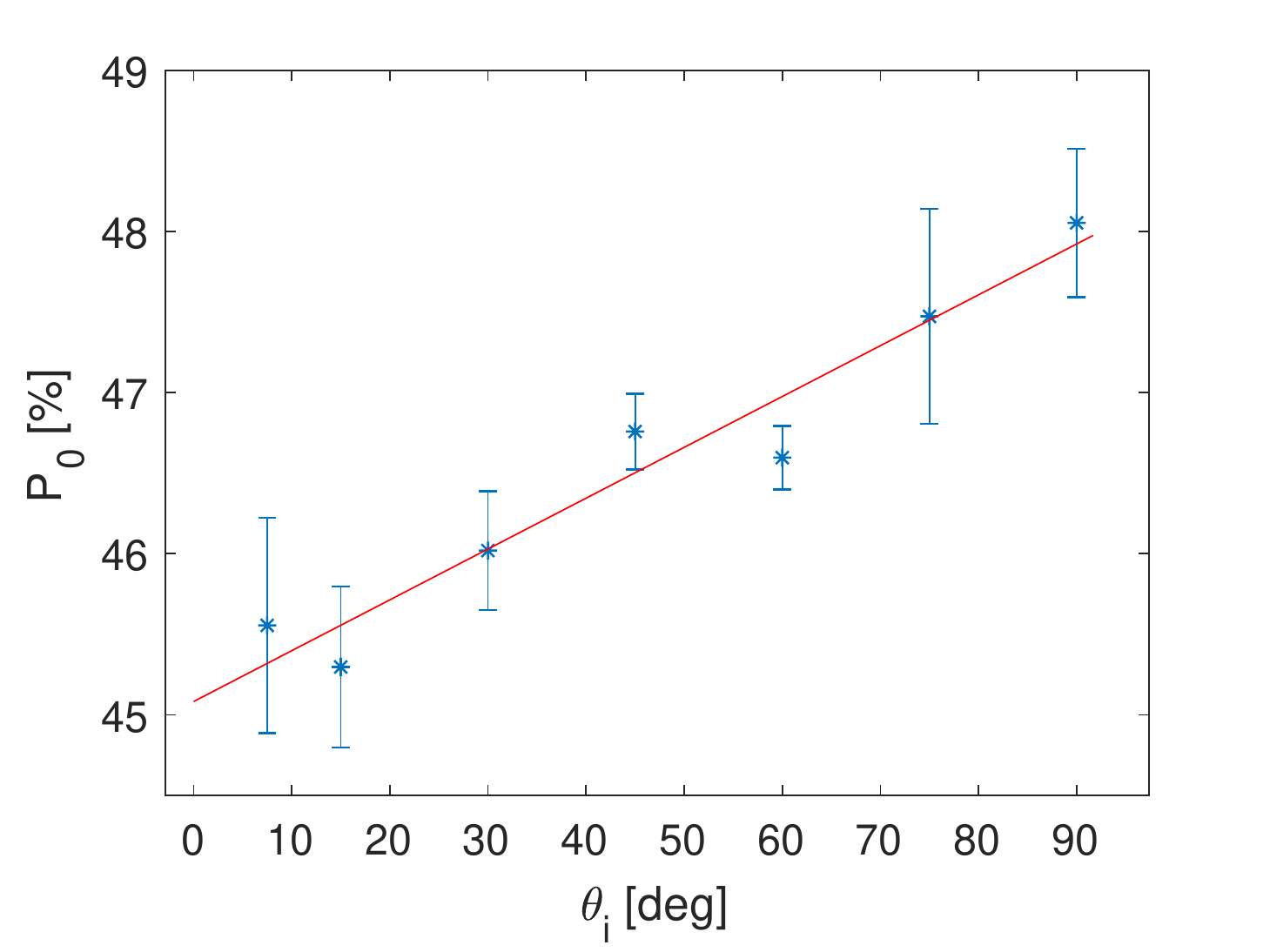}
\par\end{centering}
\caption{{\footnotesize{}The dependence of the of the $\Sigma$ and $P_{0}$
parameters on the inclination angle $\theta_{i}$ of the star-planet
system, obtained by fitting the capture probability with $P\left(v\right)=P_{0}\exp\left(-v^{2}/2\Sigma^{2}\right)$.
The error bars mark the 95\% confidence level error range. The dependence
was fitted with a simple linear function $f\left(\theta_{i}\right)=a\theta_{i}+b$
for both $\Sigma$ and $P_{0}$. The parameter $\Sigma$ is best fitted
with $a=\left(0.53\pm0.5\right)\times10^{4}\protect\units{cm\cdot s^{-1}\cdot rad^{-1}}$
and $b=\left(2.47\pm0.05\right)\times10^{4}\protect\units{cm\cdot s^{-1}}$,
while the best fit for $P_{0}$ is achieved with $a=0.018\pm0.005$
and $b=0.451\pm0.05$. The errors range is of 95\% confidence levels
as well.}}
\label{fig:The-dependence-on-the-inclination}
\end{figure}

The fitted Gaussians profiles, shown in Fig.~\ref{fig:Probability_Vs_Velocity_all_inc-1},
predict the fraction of captures $P$ for a given velocity. To account
for the dependence on the inclination angle $\theta_{i}$, we analyze
the variations of the standard deviation $\Sigma$ and the $P_{0}$
with the inclination angle, which are shown in Fig.~\ref{fig:The-dependence-on-the-inclination}.
 We fit a linear function for both $\Sigma$ and $P_{0}$, so that
the capture probability, given in c.g.s units, is approximated analytically
with $P\left(v,\theta_{i}\right)=\left(0.018\theta_{i}+0.451\right)\exp\left[-v^{2}/10^{8}\left(0.75\theta_{i}+3.5\right)^{2}\right]$.
The orbital radius of the bound planet $r_{B}$ was constant throughout
the simulations, so a possible dependence of the capture probability
on $r_{b}$ is not accounted for; however, a Jupiter-Sun separation
is indeed typical according to the NASA Exoplanet Archive\footnote{http://exoplanetarchive.ipac.caltech.edu/}.

\section{The capture rate}

We evaluate the capture rate for the galactic thin disk, which contains
most of the stars in our galaxy $\left(N_{s}\sim10^{11}\right)$.
With a moderate typical velocity dispersion of $\sigma_{v}=40\units{km\cdot s^{-1}}$
and higher metallicity values, the thin disk may be considered as
a plausible source of free-floaters and with higher capture rates
\citep{murdin2001encyclopedia}. With a relative velocity dispersion
of $\sigma_{rv}=\sqrt{2}\sigma_{v}$, we use a Boltzmann velocity
distribution function for the region in question. The interaction
cross section that results for velocities lower than $v_{\mathrm{lim}}=0.07\units{km\cdot s^{-1}}$
covers an area that is larger then the typical separation between
the stars. To overcome this difficulty, we use a constant interaction
cross section of $\mathcal{B}_{\mathrm{lim}}=n_{stars}^{-\nicefrac{2}{3}}\simeq3.7\units{pc^{2}}$
for velocities below $v_{\mathrm{lim}}$, where $n_{stars}=0.14\units{pc^{-3}}$
is the number density of stars at the solar neighborhood. This accounts
for slow free-floaters that enter the unit volume of the star, while
any free-floater outside of this volume is dominated by the gravity
of another star. After integrating with respect to velocity and solid
angle (Eq. \ref{eq:Spesific Rate}), we obtain the capture rate as
a function of the stellar mass and the planetary-mass ratio $\mathrm{R}\left(M,\rho\right)$.

We assume that a Jupiter mass is a typical mass of a free-floater,
as predicted through observations by \citet{microlensing2011unbound},
and use the probability distribution function given by \citet{malhotra2015mass}
for planetary masses. This distribution function is based on masses
of observed Kepler exoplanets for which a debiased distribution was
constructed. We also use a present-day mass function for the galactic
thin disk as presented by \citet{chabrier2003galactic} and constructed
from the observed luminosity function.

The number density of free-floaters $n_{f}$ is assumed to be constant,
as the density distribution of the free-floaters is yet unknown. Using
a number density of $n_{f}=0.24\units{pc^{-3}}$, as estimated by
\citet{microlensing2011unbound} from microlensing surveys, the capture
rate in the galactic thin disk is $R\simeq6.4\times10^{-6}\units{yr^{-1}}$
(Eq.~\ref{eq:Total Rate}), which is about one ``temporary capture''
every $1.5\times10^{5}\units{yr}$. One problem with this prediction
is that the estimated number density is model dependent, since a delta-function
mass distribution was assumed for the free-floaters population. A
second problem is that the microlensing event detection was limited
to free-floaters with masses of $m_{f}>0.1M_{J}$. \citeauthor{microlensing2011unbound}
also tested a power-law distribution for the free-floaters and received
a number density that is 11 times higher. If we adopt the ex-situ
approach for the origin of free-floater as introduced by \citet{dado2011misaligned},
then a possible number density of $n_{f}\approx200\units{pc^{-3}}$
is predicted, for which the capture rate in the galactic thin disk
is $R\approx0.0046\units{yr^{-1}}$; that is, we expect a ``temporary
capture'' to occur about every 218 years.

\begin{figure}[h]
\begin{centering}
\includegraphics[width=\columnwidth]{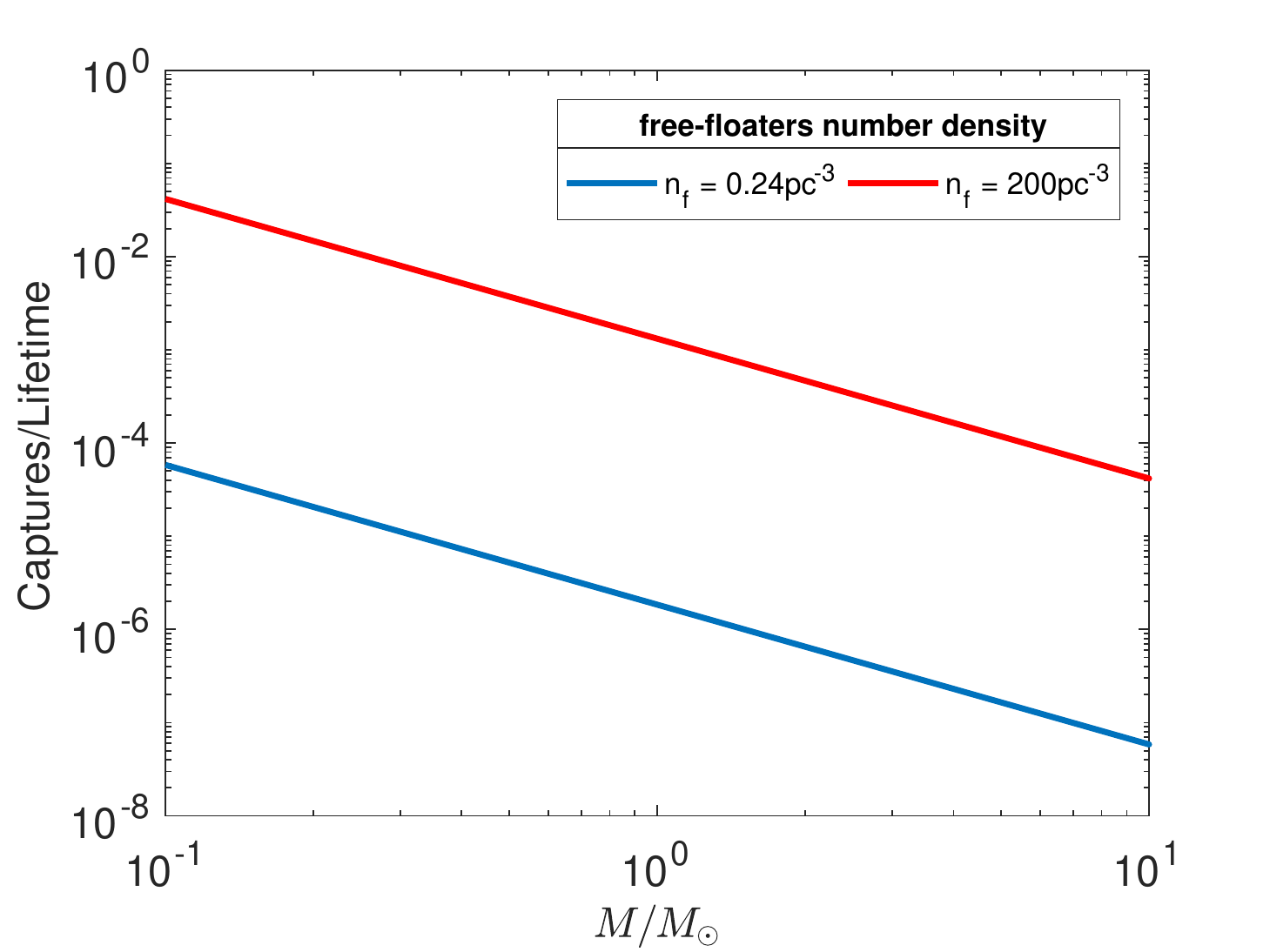}
\par\end{centering}
\caption{{\footnotesize{}Expected stellar disk captures during the lifetime
of the star as a function of the its mass. The red curve corresponds
to the predicted free-floaters number density $\left(n_{f}\right)$
assuming that their origin are explosive death of stars, while the
blue one corresponds to a free-floaters number density derived from
microlensing surveys.}}
{\footnotesize{}\label{fig:Expected-captures-during-lifetime}}{\footnotesize \par}
\end{figure}

The lifetime of stars varies according to their mass, due to fact
that the mass is the energy source of the star and it dictates the
energy output $L$ (luminosity). A rough relation between the lifetime
and mass is derived by evaluating the time required for the star to
consume itself, 
\begin{equation}
\tau\left(M\right)=\frac{fMc^{2}}{L}\sim\tau_{\odot}\left(\frac{M}{M_{\odot}}\right)^{-2.5},
\end{equation}
where $\tau_{\odot}=10^{10}\units{yr}$ is the lifetime of the Sun.
We use this estimation to evaluate the number of temporary captures
that are expected for different stellar masses. As displayed in Fig.~\ref{fig:Expected-captures-during-lifetime},
it turns out that about one out of every $\sim760$ solar-mass stars
is expected to experience capture a free-floater during its lifetime,
while for red-dwarf stars with a mass of $0.1M_{\odot}$ the expectancy
is a capture by one out of every 25 stars. 
 As more then 85\% of the stars are of sub-solar masses, they are
expected to be the main contributors to the capture rate. We estimate
the expected fraction of stars that will capture a free-floater during
their lifetime by calculating the expected value of $\mathrm{R}\left(M\right)\tau\left(M\right)$,
where we use the initial stellar mass-function. We predict that $\sim1\%$
of all $0.1M_{\odot}<M<2M_{\odot}$ stars are expected to experience
a capture during their lifetime.

\begin{figure}[h]
\begin{centering}
\includegraphics[width=\columnwidth]{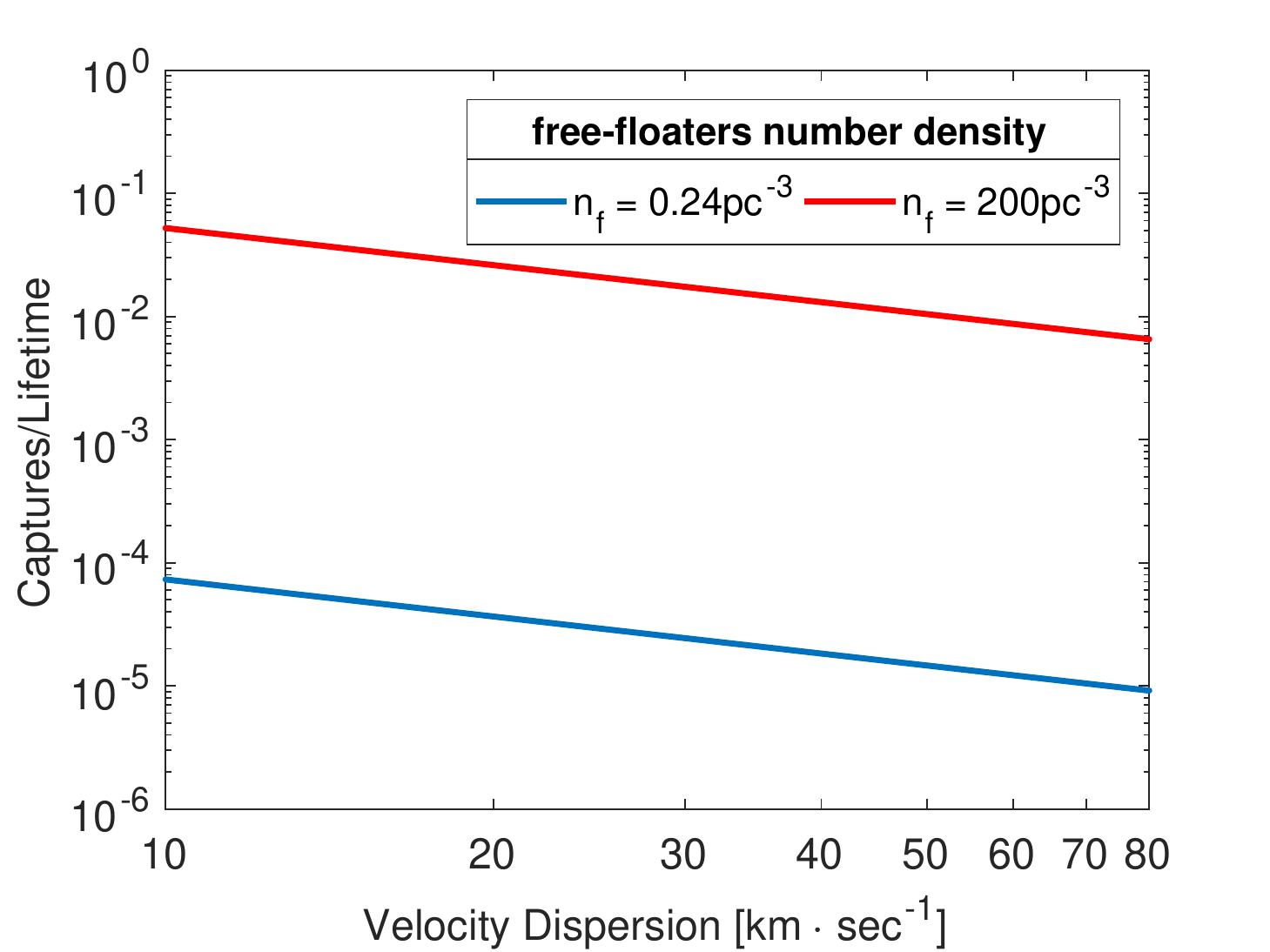}
\par\end{centering}
\caption{{\footnotesize{}Percentage of stars that are expected to capture a
free-floater during their life time as a function of the velocity
dispersion. The red curve corresponds to the predicted free-floaters
number density $\left(n_{f}\right)$ assuming that their origin are
explosive death of stars, while the blue one corresponds to a free-floaters
number density derived from microlensing surveys.}}
{\footnotesize{}\label{fig:Percentage-of-stars Vs Dispersion}}{\footnotesize \par}
\end{figure}

The capture rate has a strong dependence on the velocity dispersion,
as illustrated in Fig.~\ref{fig:Percentage-of-stars Vs Dispersion},
indicating that the majority of captures are expected to take place
in very cold regions (i.e. of slow planets). 

\section{Discussion}

The vast majority of the capture events ends up with a very small
binding energy between the free-floater and the host star, resulting
in very elongated and eccentric orbits. Captured free-floaters may
stay bound for a finite period of time until they gain energy back
from planet-planet perturbation and get ``kicked out'' back to the
interstellar medium. The following perturbations can also cause the
captured free-floater to lose energy and thus ``tighten'' its orbit.
Longer two-dimensional simulation carried by \citet{varvoglis2012interaction}
showed that 90\% of the captured free-floaters eventually got ejected
back to the interstellar medium. In most of our scattering simulations,
the bound planet experienced only one perturbation, as the equations
of motion were integrated up to the point where the free-floater travels
about twice its initial distance.

The strong velocity dependence of the cross section results in a very
low capture rate of one capture every 218 years in the galactic thin
disk, implying that captures are expected to occur in low velocity
dispersion regions. Moreover, we simulated low-velocity scatterings
because captures are expected at this velocities range. Since typical
velocities in our galaxy are much higher, scattering simulations at
higher velocities are needed to avoid extrapolations.

Moreover, 42\% of the detected planetary systems contain more then
one planet, and about one third of the stellar systems contain at
least two stars. Simulating a planetary system with multiple bound
planets should produce more boundaries between flyby and capture regions
(see Figs.~\ref{fig:Energy-map 700x500} and \ref{fig:An-outcome-map 175x125}),
at which significant orbit perturbations occur and hence increase
the fraction of strong captures.

One must remember that our simulation assumed point masses and only
the gravitational forces that they produce. We did not include any
additional effects that may subtract energy from the interacting free-floater,
such as tidal force and collision with clouds of debris; these should
increase the capture fraction and tighten the orbits of captured planets.
Not all captures, however, would be subjected to dissipation mechanisms
in the same manner. Treating the eccentricity and the semi-major axis
as additional variables of the cross section would allow us to distinguish
between captures that are too loose and inclined to remain bound,
and those that might get tighter and circularized on reasonable time-scales.\\

\end{document}